\documentstyle[12pt,aasms4]{article}
\def \yskip{\penalty-50\vskip3pt plus 3pt minus 2pt}
\def \pp{\par \yskip \noindent \hangindent .4in \hangafter 1}
\def \abc#1#2#3#4 {\pp#1, {\sl#2}, {\bf#3}, #4}
\def \blank {\lower 5pt\hbox to 0.75in{\hrulefill}}
\def \kms{km~$\rm{s}^{-1}$}

\def \c2{$\rm{cm}^{-2}$}

\def \mum{$\mu$m}
\def \lsol{L$_{\odot}$}

\newfont{\rten}{cmr10} 
\def \arcdeg{\hbox{$^\circ$}}
\def \arcmin{\hbox{$^\prime$}}
\def \arcsec{\hbox{$^{\prime\prime}$}}

\begin{document}
\normalsize
\title{The Optical Proper Motions of HH 7-11 and Cep E (HH 377)}

%%\centerline{Submitted May 28, 2001 to AJ. Not released yet}

\author{Alberto Noriega-Crespo}
\affil{SIRTF Science Center, California Institute of Technology}
\affil{IPAC 100-22, Pasadena, California, 91125}
\affil{Electronic mail: alberto@ipac.caltech.edu}

\affil{and}

\author{Peter M. Garnavich}
\affil{University of Notre Dame}
\affil{Nieuwland Science Hall 213, Indiana, 46556}
\affil{Electronic mail: pgarnavi@nd.edu} 

\begin{abstract}
A key ingredient in understanding the dynamics of stellar outflows 
is their proper motion.  We have used optical images in the [SII] 
emission at 6717/31 \AA~ and the red Digitized Palomar 
Observatory Sky Survey (DSS)
plates to determine the proper motion of HH 7-11 system and the optical 
knot of Cep E (HH 377). The DSS plate measurements span nearly 37 years
for both HH 7-11 and HH 377 and have wide field of view, which
allows an accurate determination of the proper motions despite their
relatively low angular resolution. The optical images, with higher angular
resolution, cover a shorter period of 7 and 4 years, respectively, and
have been used to complement the DSS measurements. 
From the DSS plates we have found that HH 377
has a proper motion of $0.031\pm 0.003$ arcsec/yr 
with a PA = 206\arcdeg, i.~e. moving away from IRAS 230111+63, 
that at a distance of 730 pc corresponds to a tangential velocity 
of $107\pm 14$~\kms. The values obtained from the optical images
are consistent with these measurements.
Similarly, the proper motions of HH 7 - 11 range 
from $0.015\pm 0.009$ (HH 9) to $0.044\pm 0.007$ (HH 11) arcsec/yr, 
and the flow is moving away from SVS 13 with a mean PA$\sim$ 136\arcdeg.
At a distance of 330 pc, these motions correspond to tangential 
velocities of $\sim 25 - 70$ \kms, i.~e. comparable to the original 
values obtained by Herbig \& Jones (\cite{her83}). The measurements
from the optical CCD [\ion{S}{2}] images are again consistent with these 
motions, although in detail there are some difference, particularly for
HH 7 and HH 10.

\end{abstract}

\keywords{stars: ISM: Herbig-Haro objects --- ISM: jets and outflows
 --- ISM: kinematics and dynamics --- stars: winds and outflows}

\section{Introduction}
\label{intro}

It has been nearly 20 years since the association between Herbig-Haro
(HH) objects and bipolar jets from newly formed stars was established. 
A fundamental element for this conclusion was the determination of
the proper motions of the HH 1/2 system, which showed atomic/ionic gas
moving away in opposite directions from an embedded source at flow 
velocities of $\sim 300-450$ \kms~(Herbig \& Jones \cite{her81}; 
Pravdo et al.~\cite{pra85}).
The fact that large format CCDs with small pixels have been around almost 
a decade has made it possible to measure relative proper motions more
accurately, replacing the photographic methods.
These relative proper motions measurements have been particularly 
successful for nearby Herbig-Haro outflows, where flow velocities of 
$\sim 100-400$ \kms,  can produce measurable pixel shifts within 4-5 years 
for an object at a distance of $\sim 150-450$ pc, i.e. the distances to the 
Taurus and Orion molecular clouds, respectively. The method can be applied
also on shorter time scales using high angular resolution HST images, as
has been the case for HH 30 (Burrows et al.~\cite{burr96}) and 
HH 111 (Hartigan et al.~\cite{hart01}).

The results obtained by this method for objects such as 
HH 1/2 (Eisl\"offel, Mundt \& B\"ohm~\cite{eis94b}),
HH 32 (Curiel et al.~\cite{curi97}), 
HH 34 (Eisl\"offel \& Mundt ~\cite{eis92}; 
Heathcote \& Reipurth ~\cite{heat92}; Devine et al.~\cite{dev97}),
HH 46-47 (Eisl\"offel \& Mundt \cite{eis94a}) and HH 110/111 
(Reipurth, Raga \& Heathcote~\cite{reip96}; Reipurth, Raga \& 
Heathcote~\cite{reip92}) among others, have provided a unique picture of 
the dynamical behavior of these outflows. High angular resolution radio 
observations based on the same principle, have confirmed some of 
these results (e.~g. for HH 1/2, Rodr\'{\i}guez et al. 1990) and given new 
measurements for systems such as HH 80-81 (Marti, Rodr\'{\i}guez \& 
Reipurth \cite{mar98}). 
The method is now being used in the near infrared, thanks to 
large format infrared arrays that allow to cover wider fields of view and 
include more reference stars. Objects like HH 1/2 (Noriega-Crespo et al.~
\cite{nori97}, HH 46/47 (Micono et al.~\cite{mico98}), GGD 37 
(Raines et al.~\cite{rain00}), HH 7-11 and 25/26 
(Chrysostomou et al.~\cite{chry00}) and 
OMC-1 (Lee \& Burton~\cite{lee00}),
have proper motions measured either in H$_2$ 2.12\mum~or [Fe~II] 1.67\mum.
Nevertheless we believe that in some cases the smaller field of view and 
shorter time baseline provided by the IR arrays, make some of these results 
less conclusive than optical measurements.

The present work is motivated precisely by the difference between 
the optical and the near infrared measurements of the proper motions of 
HH 7-11. The tangential flow velocities of the atomic/ionic gas inferred 
by using the photographic plate method are $\sim 20 - 60$ 
\kms~(Herbig \& Jones~\cite{her83}), 
while those of the molecular gas (H$_2$) determined by using 
the near infrared imaging method are $\sim 300-400$ \kms 
(Chrysostomou et al.~\cite{chry00}). 
This difference is difficult to explain from the theoretical
point of view, since in other systems we observe quite similar 
velocities for both gas components (Noriega-Crespo et al.~\cite{nori97};
Micono et al.~\cite{mico98}). 
In the case of HH 7-11, the driving source SVS 13 is a relatively 
low mass proto-stellar object, that makes it  even harder to explain 
the large H$_2$ proper motions (see $S$4).

Another goal of this project is to measure the proper motions of 
semi-embedded outflows. This is the case for HH 7-11, where the 
counter-flow is nearly invisible, and Cep E (HH 377), where most of the 
outflow is invisible at optical wavelengths (Ayala et al.~\cite{aya00}).

In this study we determine the proper motions of HH 7-11 and Cep E (HH 377),
using digitized plates and optical CCD images.
Both outflows are clearly seen in the 
Digitized Palomar Observatory Sky Survey\footnote{The Digitized Sky Surveys
were produced at the Space Telescope Science Institute under U.S. Government
grant NAG W-2166. The images of these surveys are based on photographic
data obtained using the Oschin Schmidt Telescope on Palomar 
Mountain and the UK Schmidt Telescope. The plates were processed into 
the present compressed digital form with the
permission of these institutions.}  (DSS), first and second 
generation, and they provide a time baseline of nearly 37 
years. The CCD images cover a  period of 7 and 4 years respectively, 
with the most recent images taken in Nov. 30, 2000 (see below).

\section{Observations} 
\label{observ}

The optical images were obtained with the Fred L. Whipple (FLWO)
1.2 m telescope and AndyCam CCD camera in 1993 on Oct. 12 (HH 7-11) 
and 1996 on Sep. 16 (Cep E). A narrow band filter centered at 672.5 nm 
and a width of 3.0 nm (FWHM) was used to isolate the [SII] emission. 
Second epoch images were taken
with the 1.8 m Vatican Advanced Technology Telescope (VATT) and
VATTCCD camera on 2000 Nov. 30. A similar narrow filter centered
at 672.3 nm and 3.5 nm width was employed at the VATT. In all cases
the CCD was binned by 2 in both directions providing an effective
scale of 0.63\arcsec~per pixel at FLWO and 0.40\arcsec~per pixel 
at the VATT.

The digitized images were downloaded from the STSCI Digitized Sky Survey
(http://stdatu.stsci.edu/dss/). We found that both HH 7-11 and Cep E 
(HH 377) appear in the First Generation Survey, as well as in the Second 
Generation (Red) Survey. 
The details of the digitized plates and the ground based
observations are presented in Table 1.

\section{Analysis}
\label{analysis}

One needs to be careful when comparing photographic plates with CCD images.
We selected the DSS plates as primary reference because of their 
long time baseline that allows a more accurate measurement of the
small pixel shifts produced by the proper motions. The CCD images by
themselves cover  only a period of 7 (HH 7-11) and 4 (HH 377) years
respectively, and so we use them as secondary indicators. The idea is to
set the measurements of the CCD images in the same scale of the plates, 
and then check for consistency and/or differences between them.
 The reason to proceed in this way is to minimize systematic effects that
could arise from comparing the different bandpasses used in the plates 
and the CCD images. The digitized plates are sensitive to radiation
similar to that passing through a broad band R-filter; and so for
shock excited regions, like those in outflows, this 
means that they might include emission from collisionally excited lines 
such as [\ion{N}{2}] and [\ion{O}{1}] (so not only [\ion{S}{2}]) as well as 
H$\alpha$ recombination emission, that could spatially arise from
very different regions.

\affil{See Figure 1}
\affil{See Figure 2}

Thus all the CCD images and the second epoch digitized plates were mapped
into the DSS I plate scale of 1.7\arcsec~pixel and brought into the same 
reference system by using 15  stars common to all the images for HH 7-11 
and 10 for Cep E, respectively. 
This transformation was performed using the IRAF tasks GEOMAP 
and GEOTRAN, which take into account relative translations, rotations, 
and magnifications between images. 
The selected reference stars do not show any indication of systematic 
motions, i.~e. large radial velocities or proper motions, according to 
the data derive from SIMBAD database and other sources (e.~g. Aspin, 
Sandell \& Russell~\cite{asp94}; Strom, Vrba \& Strom~\cite{str76}). 
The selected stars for HH 7-11 and Cep E (HH 377) are
shown in Figures 1 and 2, respectively.
As in our previous projects (Lop\'ez et al.~\cite{lop98};
Noriega-Crespo et al.~\cite{nori97}; Reipurth et al.~\cite{reip93}) the 
proper motions were obtained by using a cross-correlation technique over 
small sections of a pair of images, and checked again by measuring the 
difference between centroids determined by fitting two-dimensional 
Gaussians.
The uncertainty in the measurement of well-resolved bright knots is
$\sim 0.1$ pixels, in the pixel scale of the reference frame.
Although the cross-correlation technique is quite reliable, we are limited
by the available knots/structure of the earliest DSS digitized images. 
Because of this we concentrated only on the brightest knots of the HH 7-11 
system, despite the fact that the most recent CCD images show a more
rich and complex structure. In the case of Cep E despite his
complex structure in the near infrared, only a knot from the south
lobe is optically visible (Noriega-Crespo~\cite{nori97}; 
Devine, Reipurth \& Bally~\cite{dev97}).

The difference between pixels as a function of epoch for the HH 7-11 knots,
and Cep E (HH 377) are shown in Figures 3 and 5, respectively. The 
fit to these data are  presented in Table 2, and correspond to that
of the filled squares (Figs 3 and 5).

\section{Results \& Discussion}

\subsection {HH 7-11 System}

The proper motions of the HH 7-11 outflow range from 0.015 $\pm$ 0.009 
\arcsec/yr~(HH 9) to 0.044 $\pm$ 0.007 \arcsec/yr~(HH 11), 
that at a distance of 330 pc correspond to a range of tangential 
velocities of $\sim 26 - 73$\kms~
(Table 2). Except for HH 9 which has large uncertainties and may be almost
stationary, the other objects are moving away from SVS 13,
the driving source (Figure 4), following its characteristic arc-shaped 
morphology. Overall these motions are not that different from those 
presented by Herbig \& Jones (1983) in their Figure 4. 
The total velocities can be estimated using the published 
radial velocities (Solf \& B\"ohm~\cite{solf87}),
with the largest value of $\sim 185$ \kms~for HH 11, the knot closer to the
SVS 13 source. HH 9 has not published radial velocity, so a lower limit is 
set by its proper motion. 

\affil{See Figure 3}
\affil{See Figure 4}

The DSS proper motions are 5-14 times {\it lower} than those measured 
using molecular Hydrogen 2.12 \mum~images (Chrysostomou et al.~\cite{chry00}). 
For example, compared that of HH 7 of 0.41\arcsec/yr~in H$_2$ with the
0.024\arcsec/yr~value (but see below). Needless to say that
if the atomic/ionic gas share a similar motion with the H$_2$ gas, then
it would be very easy to detect so large shifts in the optical images.
We don't believe this difference is real. We trust the optical results more
than those from H$_2$ because of the longer time-span (nearly 37 yrs 
compared with 4-5 yrs) and the larger number of well selected and measured
reference stars. These two ingredients are necessary to avoid systematic
offsets, magnifications and rotations that could bias the results in an 
unexpected fashion. 

Although the shifts measured in the  CCD [\ion{S}{2}] 
images are consistent with the ``predicted'' by the DSS proper motions, 
in detail there are some differences, particularly for HH 7 and HH 10,
as shown by the fits (dotted line) to their measurements (open circles)
in Figure 3. The proper motions for HH 7 based on the 
CCD [\ion{S}{2}] images are PM$_x$ = -0.12$\pm 0.02$\arcsec/yr  and 
PM$_y$ = 0.08$\pm 0.02$ \arcsec/yr  , which corresponds to velocities of 
V$_x$ = 190$\pm$20 \kms~ and V$_y$ = 130$\pm$20 \kms for a distance of
330pc. The tangential velocity derived from them 
is V$_{tan}$ = 230 $\pm$ 28 \kms~
with a position angle PA = 56\arcdeg$\pm$ 7\arcdeg. This velocity is
5 times faster than the mean motion obtained from the DSS plates, but it
is a bit misleading since it is based {\it only} on the bowshock, without
the Mach Disk (which do not display a shift at all), 
and it is biased by the lower angular resolution of the 1993 image.
But even if this estimate were correct, is still a factor 2 less than
that obtained for HH 7A in H$_2$ of $450\pm 44$ \kms~by 
Chrysostomou et al. 2000, that assumes a distance of 220pc.
The case of HH 10 is less dramatic, but shows the difficulty of using
only two epochs separated by a relatively short period of time, since
its shift in the Y-direction has a positive slope 
(Fig 3, open circles). 
This means going from PM$_y$ = -0.019$\pm 0.008$ \arcsec/yr
to PM$_y$ = 0.015$\pm 0.02$\arcsec/yr , with a change of position angle of 
nearly 80\arcdeg; the total tangential velocity is essentially the same
as before.

Astrophysically there are at least two reasons which indicate that the
smaller proper motions are more appropriated; one is the luminosity of the
source and the other the low excitation of the objects.
The recent interferometric maps of the HH 7-1 outflow
(Bachiller et al.~\cite{ba00}), as well as the near infrared NICMOS images 
(Noriega-Crespo et al.~\cite{nori01}), 
have confirmed that a jet arises from SVS~13 and that the molecular CO gas 
follows the path of HH 7-11 optical knot; so there is little doubt that 
SVS~13 is the outflow source. SVS~13 is a relatively low mass young stellar
 object, with a bolometric luminosity of $\sim 85$ \lsol~(Molinari, Liseau 
\& Lorenzetti~\cite{mol93}), and so we do not expect outrageously high 
outflow velocities as seen in intermediate or high protostellar mass 
objects, like e.g. HH 80-81 where flow velocities of $\sim 600-1400$ \kms~
have been measured (Marti et al.~\cite{marti93}) or Z~CMa $\ge 600$ \kms~
(Poetzel, Mundt, \& Ray~\cite{poe89})
with luminosities of $2\times 10^4$ and 3500 \lsol~(Marti et al.
~\cite{marti93}; Hartmann et al.~\cite{har89}).
The other indicator are the shock velocities themselves, since at least
at first approximation we expect high shock velocities for high flow 
velocities at the leading working surface.  Again this is what is 
observed in systems like HH 1/2 or HH 80/81, where the [O~III] emission 
confirms their higher excitation and shocks higher than 100 \kms. 
We do not see such high excitation
in HH 7-11. The shock velocities obtained from the optical spectra for 
HH 7-11 are $\sim 40$ \kms~(Solf \& B\"ohm \cite{solf87}).
The ISO observations in the mid/far infrared also set a limit of 
$\sim 40-50$ \kms~ for the shock velocities of both outflow lobes 
(Molinari et al.~\cite{mol00}), in agreement with the value derived using 
optical spectroscopic observations and consistent with the low excitation 
nature of these objects.

Finally, the optical and H$_2$ measurements are really sampling different 
gases in the case of HH 7-11, and this quite apparent from the 
superposition of 2.12 \mum~and [S~II] emission. The extreme example 
corresponds to HH 11, that appears as a bullet even in the high angular 
resolution WFPC2 images (Noriega-Crespo et al.~\cite{nori01}), while the 
H$_2$ emission is fuzzy, wide open and a few arc seconds downstream, 
as what one would expect from emission arising at the wings of a bowshock.

\subsection{Cep E - HH 377}
For HH 377 we proceed as with HH 7-11, we use the DSS plates 
to determine the mean proper motions and then we analyze the shifts
measured in the CCD images with the predicted values.
The case for HH 377 is more simple than for HH 7-11,
since we are dealing with a single optical knot.
Our measurement indicate that HH 377 is moving away from the 
IRAS 23011+6126 source (PA = 206\arcdeg$\pm$8\arcdeg)
with a proper motion of 0.031 $\pm$ 0.004 \arcsec/yr (Table 2; Fig 5), 
that at a distance of 730pc translates into a tangential velocity of 
$\sim 107$ \kms (Figure 6).
The CCD measurements are consistent with their predicted values, although
they could be higher, as indicated by the fit (dotted line) to the open squares
in Figure 5, i.e. PM$_x$ = -0.068$\pm$ 0.03 \arcsec/year and 
PM$_y$ = -0.077$\pm$ 0.03 \arcsec/year. This corresponds to V$_{tan}$ = 230
$\pm$ 98 \kms~and PA = 220\arcdeg$\pm$ 20\arcdeg. We can estimate the flow
velocity of Cep E, by combining the proper motions and the radial velocity
of HH 377. Thus, if we take the DSS mean proper motion and a
mean radial velocity of $\sim -70$ \kms~obtained from H$\alpha$, 
[SII] 6717/31 and [O~I] 6300/63 emission lines (Ayala et al. \cite{aya00}),
which leads to a total velocity of $\sim 130$ \kms. 
This flow velocity is quite 
similar to the 125 \kms obtained from the radio measurements of the 
$^{12}$CO(2-1) molecular gas emission (Eisl\"offel et al.~\cite{eis96})
for the blue wing of the outflow.

For Cep E (HH 377) the shock velocities range from $\sim 15-35$ \kms~, 
based on optical and near/mid/far infrared spectroscopic observations 
(Ladd \& Hodapp ~\cite{ladd97}; Ayala et al.~\cite{aya00}; Moro-Martin et 
al.~\cite{ama01}). The source itself, IRAS 23011+6126, a Class 0/I 
protostellar object has a bolometric luminosity of 30\lsol, i.e. a low mass 
YSO (Moro-Martin et al.~\cite{ama01}). So the power of the source and the 
magnitude of the flow and shock velocities are in agreement.

\section{Conclusions}

We have obtained the proper motions of two semi-embedded young stellar 
outflows, HH 7-11 and HH 377 (Cep E) using optical images in [S~II] and 
DSS plates epoch I and II. The HH 7-11 system has proper motions
comparable in magnitude and direction as those previously obtained
by Herbig \& Jones (1983), but almost a factor 10 smaller than those 
measured for the molecular Hydrogen gas using H$_2$ 2.12\mum~near infrared 
images.

Astrophysically is difficult to see why there should be such a large 
discrepancy, and so we have argued on favor of the optical measurements based 
on a longer time baseline and a more accurate reference frame 
between the images at different epochs. 
If the optical measurements are correct, then 
this questions theoretical scenarios based on speeds exceeding 400 \kms~
for HH 7. Presently, however, we can not rule out ideas that explain the 
observed excitation in HH 7/11 as the result of a fast jet propagating 
through a stationary medium which contains several dense clumps or 
structures (Chrysostomou et al.~\cite{chry00}).
In this context it interesting to notice that HH 9, given its proper motion,
may not be part of the HH 7-11 outflow, although
 certainly it is a shock excited 
condensation. We should point out, however, that ASR 57 another shock 
excited clump $\sim 90$\arcsec~south-east from HH 7 seems
to be related to the HH 7-11 outflow (Aspin, Sandell \& Russell
~\cite{asp94}), and might be the outcome of a previous major ejection event
from SVS 13, and if so, we are not dealing necessarily with a 
stationary medium.
 
The case for HH 377 (Cep E) is more simple, and its proper motion is 
quite consistent with other velocity indicators, e.~g. the  $^{12}CO(2-1)$
outflow velocity wings, and with the overall bipolar morphology driven
by IRAS 23011+6126.  

\affil{See Figure 5}
\affil{See Figure 6}

\acknowledgments

We thank S. Curiel for making available to us his 1996 HH 7-11 optical 
images, to A. Moro-Martin for her careful reading of the manuscript, and
last but not least, to the referee Jochen Eisl\"offel for his insightful
comments.
The Digitized Sky Surveys were produced at the Space Telescope 
Science Institute under U.S. Government grant NAGW-2166. The images of 
these surveys are based on photographic data obtained using the Oschin 
Schmidt Telescope on Palomar Mountain. The National Geographic Society - 
Palomar Observatory Sky Atlas (DSS-I) and the Second Palomar Observatory 
Sky Survey (DSS-II) were made by the California Institute of Technology.

{}

\makeatletter
\def\jnl@aj{AJ}
\ifx\revtex@jnl\jnl@aj\let\tablebreak=\nl\fi
\makeatother

\begin{deluxetable}{rlcl}
\tablewidth{14cm}
\tablecaption{HH 7-11 and Cep E (HH 377) Data\label{tbl-1}}
\tablehead{
\colhead{Date} & \colhead{Band} & \colhead{Scale~(\arcsec/pixel)} 
& \colhead{Telescope}}
\startdata
            HH 7-11  &  & \nl 
10/24/1955 & DSS E Red  & 1.70 & Palomar 48in Schmidt\nl
09/30/1992 & DSS II Red IIIaF & 1.00 & Palomar 48in Schmidt\nl
10/12/1993 & [SII] 6717/6731 \AA~ & 0.65 & Mt. Hopkins 1.2m\nl
11/29/2000 & [SII] 6717/6731 \AA~ & 0.40 & VATT\nl
  Cep E & &\nl
10/31/1953 & DSS E Red & 1.70 & Palomar 48in Schmidt\nl
09/03/1991 & DSS II Red IIIaF & 1.00 & Palomar 48in Schmidt\nl
09/16/1996 & [SII] 6717/6731 \AA~ & 0.65 & Mt. Hopkins 1.2m\nl
11/30/2000 & [SII] 6717/6731 \AA~ & 0.40 & VATT\nl
\enddata
\end{deluxetable}

\clearpage

\makeatletter
\def\jnl@aj{AJ}
\ifx\revtex@jnl\jnl@aj\let\tablebreak=\nl\fi
\makeatother

\begin{deluxetable}{lrccll}
\tablewidth{15cm}
\tablecaption{HH 7-11 and Cep E (HH 377) Proper Motions\label{tbl-2}}
\tablehead{
\colhead{Object} & \colhead{PM$_x$(\arcsec/yr)} & \colhead{PM$_y$(\arcsec/yr)} 
& \colhead{PM$_{tot}$(\arcsec/yr)} & \colhead{PA(\arcdeg) ~~V$_{tan}$(\kms)\tablenotemark{a}}} 
\startdata
HH~7  & $0.022\pm 0.014$ & $-0.007\pm 0.003$ &  $0.024\pm 0.014$ & $108\pm 10$ ~~$38\pm 23$ \nl
HH~8  & $0.008\pm 0.001$ & $-0.017\pm 0.005$ &  $0.019\pm 0.005$ & $155\pm 21$ ~~$30\pm 9$ \nl
HH~9  &$-0.010\pm 0.006$ &$~~0.011\pm 0.006$ &  $0.015\pm 0.009$ & $318\pm 31$ ~~$26\pm 15$ \nl
HH~10 & $0.010\pm 0.005$ & $-0.019\pm 0.008$ &  $0.022\pm 0.010$ & $152\pm 50$ ~~$36\pm 14$ \nl
HH~11 & $0.039\pm 0.005$ & $-0.020\pm 0.006$ &  $0.044\pm 0.007$ & $117\pm ~6$ ~~~$73\pm 12$ \nl
      &                  &                   &                   &               \nl
HH 377&$-0.013\pm 0.004$ & $-0.028\pm 0.006$ &  $0.031\pm 0.004$ &  $206\pm ~8$ ~~$107\pm 14$ \nl
\enddata
\tablenotetext{a}{For distances: d(HH 7-11) = 330 pc \& d(Cep E) = 730 pc}
\end{deluxetable}

\clearpage

\begin{deluxetable}{lrlll}
\tablewidth{15cm}
\tablecaption{HH 7-11 and Cep E (HH 377) Velocities\label{tbl-3}}
\tablehead{
\colhead{Object} & \colhead{V$_{rad}$\tablenotemark{a,b}} & \colhead{V$_{tan}$}
& \colhead{V$_{total}$} & \colhead{$\gamma$\tablenotemark{c}}\\
\colhead{} & \colhead{(\kms)} & \colhead{(\kms)} & \colhead{(\kms)} 
& \colhead{(\arcdeg)}}
\startdata
HH~7  & $-51\pm 20$  & $38\pm 23$  & $63\pm 30$  & $-53\pm 20$ \nl
HH~8  & $-57\pm 20$  & $30\pm 9$   & $64\pm 15$  & $-62\pm 13$ \nl
HH~9  & \nodata      & $26\pm 15$  & \nodata     & \nodata \nl
HH~10 & $-35\pm 20$  & $36\pm 14$  & $50\pm 24$  & $-44\pm 19$ \nl
HH~11 & $-175\pm 20$ & $73\pm 12$  & $190\pm 19$ & $-67\pm~~6$  \nl
      &              &             &             &    \nl
HH 377& $-70\pm 10$  & $107\pm 14$ & $128\pm 12$ & $-33\pm~~5$ \nl
\enddata
\tablenotetext{a}{From Solf and B\"ohm 1987 for HH 7, 8, 10 and 11}
\tablenotetext{b}{From Ayala et al. 2000 for HH 337}
\tablenotetext{c}{From $\gamma$ = tan$^{-1}$(V$_{rad}$/V$_{tan}$)}
\end{deluxetable}

\clearpage
\begin{center}
Figure Captions
\end{center}

\figcaption[anc_f1.ps]{References stars for the HH 7-11 outflow
in the DSS Second Generation image, J2000 coordinates and FOV = 7\arcmin.
\label{f1}}

\figcaption[anc_f2.ps]{References stars for the Cep E (HH 377)
outflow in the DSS Second Generation image, J2000 coordinates and 
FOV = 7\arcmin. HH 377 lies in between reference stars 1 and 2.
\label{f2}}

\figcaption[anc_f3.ps]{The difference in pixels X-direction (left)
and Y-direction (right) for the HH 7-11 images as a function of time (yrs).
The filled squares correspond to the DSS measurements, while the open
circles to those from [\ion{S}{2}].
\label{f3}}

\figcaption[anc_f4.ps]{A graphic display of the tangential velocities of 
HH 7-11, based on the proper motion measurements.
\label{f4}}

\figcaption[anc_f5.ps]{The difference in pixels (X and Y directions) 
as a function of time (yrs) for the Cep E (HH 377) images.
The filled squares correspond to the DSS measurements, while the open
circles to those from [\ion{S}{2}].
\label{f5}}

\figcaption[anc_f6.ps]{A graphic display of the tangential velocities of Cep E
(HH 377), based on the proper motion measurements.
\label{f6}}

%\clearpage

%\begin{figure}[v]
%\plotone{anc_f01.ps}
%\end{figure}

%\clearpage

%\begin{figure}[v]
%\plotone{anc_f02.ps}
%\end{figure}

%\clearpage

%\begin{figure}[v]
%\plotone{anc_f03.ps}
%\end{figure}

%\clearpage

%\begin{figure}[v]
%\plotone{anc_f04.ps}
%\end{figure}

%\clearpage

%\begin{figure}[v]
%\plotone{anc_f05.ps}
%\end{figure}

%\clearpage

%\begin{figure}[v]
%\plotone{anc_f06.ps}
%\end{figure}

%\clearpage

\end{document}